\documentclass[aps,twocolumn,prl]{revtex4}
\usepackage[utf8]{inputenc}
\usepackage{graphicx}
\usepackage{latexsym}
\usepackage{amsmath}
\usepackage{amssymb}
\usepackage{psfrag}
\usepackage{color}
\usepackage{bm}
\newcommand{\beq}{\begin{equation}}
\newcommand{\eeq}{\end{equation}}
\newcommand{\ba}{\begin{eqnarray}}
\newcommand{\ea}{\end{eqnarray}}
\newcommand{\bee}{\begin{eqnarray}}
\newcommand{\eee}{\end{eqnarray}}
\newcommand{\bc}{\begin{center}}
\newcommand{\ec}{\end{center}}

\graphicspath{{figures/}}
\makeatletter
\begin{document}

\title{Stable Configurations Of 
Charged Sedimenting Particles}
\author{C I Trombley \& M L Ekiel-Je\.zewska}
\affiliation{ Institute of Fundamental Technological Research, Polish Academy of Sciences, Pawi\'nskiego 5b, 02-106 Warsaw, Poland}
\date{\today}
\begin{abstract}
The qualitative behavior of charged particles in a vacuum is given by Earnshaw's Theorem which states that there is no steady configuration of charged particles in a vacuum which is asymptotically stable to perturbations. In a viscous fluid, examples of stationary configurations of sedimenting uncharged particles are known, but they are unstable or neutrally stable - they are not attractors. In this paper, it is shown by example that two charged particles settling in a fluid may have a configuration which is asymptotically stable to perturbations, for a wide range of charges, radii and densities. The existence of such ``bound states'' is essential from a fundamental point of view and it can be significant for dilute charged particulate systems in various biological, medical and industrial contexts. 
\end{abstract}
\maketitle
Earnshaw's Theorem gives fundamental insights into 
the stability of charged systems. Introduced in \cite{earnshaw1839}, the theorem states that there is no stable equilibrium of charged particles distributed in a vacuum without boundary. An informal reading is that electrostatic interactions are inherently destabilizing and one must add e.g. boundaries or stabilizing forces \cite{levin2005}. Historically, Earnshaw's Theorem informed the development of models of the stability of matter and studies of qualitative features of charged systems \cite{levin2005,jones1980}. Finding the stable configurations allowed by Earnshaw's Theorem when a spherical boundary is imposed - the "Thompson problem" - is an active field \cite{bondarenko2015}. Earnshaw's theorem underpins classical models of Wigner crystallization (for instance, see \cite{nazmitdinov2017}). It even allows one to find quantitative limits on parameters for stable classical models of complex molecules \cite{mohammad2004}. In this Letter, we  show that the presence of an unbounded electrically neutral fluid can stabilize 
systems of charged 
microparticles. 

At micro and nano scales, both active "agents" and passive objects, whether biological \cite{kantsler2012,kang2013,liu2018morphological} or inorganic materials \cite{perazzo2017,pawlowska2017lateral}, and naturally or artificially made, have been modeled theoretically as 
particles in a fluid. In general, such particles can have complex shapes and be deformable. Their rich dynamics 
have been extensively investigated \cite{smith1999,becker2001,lagomarsino2005hydrodynamic,young2007,harasim2013,farutin2016dynamics,gruziel2018,bukowicki2018different}.  
The development of 
microfluidics, Lab-On-Chip technologies \cite{nunes2012}, advances in medicine and the design of innovative materials and devices - e.g.. to carry drugs \cite{edwards1997} or treat wastewater \cite{coutinho2009} - depends on 
this research. 


The concept of a non-inertial "Stokes flow", introduced in \cite{stokes1851}, holds a central place in the theory of the dynamics of micro and nano particles  \cite{batchelor1967,kim2005}. In particular, Stokes equations are widely used to determine the influence of a viscous fluid on the dynamics of particles experiencing external forces, such as gravity or in a centrifuge \cite{lagomarsino2005hydrodynamic,caflisch1988,lecoq1993,ekiel2014class,bargiel2014,saggiorato2015}. For a single particle, Stokes flow is an appropriate model when the particle has reached its terminal velocity, its so-called "Stokes velocity". The terminal velocity is reached swiftly at a microscale. 
In systems of microparticles in a Stokes flow, the velocity of each particle is a linear combination of the forces on every particle. The coefficients of this combination depend on positions of all the particles. 

The goal of this Letter is to find 
asymptotically stable configurations of two sedimenting charged particles. The existence of such "attractive states" (configurations such that if the particles were disturbed from this configuration then they would tend to return) may be of a great significance for sedimenting suspensions which 
exhibit electrostatic interparticle interactions.

First, we briefly outline known results for uncharged particles. Owing to reversibility of Stokes equations, identical spherical sedimenting particles can form steady configurations, such as e.g. horizontal regular polygons made of arbitrary numbers of particles \cite{hocking1964}. More simply, any arrangement of two identical particles in free space is steady. However, these steady configurations are at most neutrally stable therefore are not attractive. 

The more interesting case of two spherical uncharged sedimenting particles with different radii and densities was examined in the seminal paper \cite{wacholder1974}. If the particles are far enough from each other that their interaction can be neglected, then particle A with a larger Stokes velocity will fall faster than particle B with a smaller Stokes velocity. Intuition may suggest that if particle A is above particle B with their centers in a vertical line, then they will tend to approach each other no matter what their distance. Counterintutively, it was found that, in a certain range of parameters, the particles do not tend to touch each other (in an infinite time) but instead "capture" each other at a distance a bit larger than the sum of their radii \cite{wacholder1974}. Even more surprisingly, particle A can move slower than B if the inter-particle distance is smaller than its steady value. In this uncharged system, vertical steady configurations are stable against vertical but unstable with respect to horizontal perturbations.

The main idea of this Letter is to introduce charge to such a system, to find a steady vertical configuration 
and check
if electrostatic attraction between the particles will cause them to 
come back to 
the steady configuration if perturbed. In the following, we will show that indeed this is the case -- we discover stable configurations. Counterintuitively to Earnshaw's Theorem in vacuum, electrostatic interactions between charged particles in fluids can play a stabilizing role. 

	We now introduce a model of two charged, spherical particles settling under gravity in a fluid of dynamic viscosity $\mu$. We assume that Brownian motion, fluid compressibility and inertia are irrelevant and we describe the fluid flow by the 
	Stokes equations \cite{batchelor1967,kim2005}. 
	Thus, the external forces on the 
	particles 
	are in balance with the fluid resistance forces, and therefore the dynamics of particles is described by a system of 
	first order differential equations. 

	We denote particle radii by $a_1$ and $a_2$. Let $M_1$ and $M_2$ represent the mass of particle 1 and 2. The reduced density of each particle is the difference between its density and the density $\rho$ of the 
	fluid. Similarly, $m_1 \!= \!M_1\! - \!\frac{4}{3}\pi a_1^3 \rho$ and $m_2 \!= \!M_2\! -\! \frac{4}{3}\pi a_2^3 \rho$ are the reduced masses. We assume $m_2 > 0$ with other cases covered in Supplemental Material. Let $\mathbf{r}_1$ and $\mathbf{r}_2$  be the positions of the centers of particle 1 and 2. Then the relative position is $\mathbf{d} \!=\! \mathbf{r}_2\!-\!\mathbf{r}_1$. 
	We choose a coordinate system so that the particle centers and the direction of gravity are  in the plane $y\!=\!0$ and $\hat{\mathbf{z}}$ is a unit vector pointing anti-parallel to 
	the constant gravitational 
	field $\bm{g}$. We can now write the 
	superposition of electrostatic and gravitational 
	forces on the particles 1 and 2 
	\begin{eqnarray}
		\mathbf{f}_1 &=& - k q_1 q_2 \frac{\mathbf{d}}{| \mathbf{d} |^3} - m_1 g \hat{\mathbf{z}}\\
		\mathbf{f}_2 &=& k q_1 q_2 \frac{\mathbf{d}}{| \mathbf{d} |^3} - m_2 g \hat{\mathbf{z}}
	\end{eqnarray}
	\noindent where $k$ is Coulomb's constant, $q_1$ and $q_2$ are the charges on particles 1 and 2, 
	 $|\mathbf{v} |$ is the length of any vector $\mathbf{v}$, and $g=|\bm{g}|$. 
	Assuming point-like charges, we consistently take a point particle approximation for the interaction with the fluid \cite{kim2005}, and 
	we obtain the following system of differential equations:
	\begin{eqnarray}
		\dot{\mathbf{r}}_1 &=& \frac{1}{8\pi\mu} \bm{G} \cdot \mathbf{f}_2 + \frac{1}{6\pi\mu a_1} \mathbf{f}_1 \label{r1} \\
		\dot{\mathbf{r}}_2 &=& \frac{1}{8\pi\mu} \bm{G} \cdot \mathbf{f}_1 + \frac{1}{6\pi\mu a_2} \mathbf{f}_2
		\label{r2}
	\end{eqnarray}
	\noindent where $G_{ij} = \delta_{ij} / | \mathbf{d} | + d_i d_j / | \mathbf{d} |^3 $ is the Green's tensor for Stokes' equations in an unbounded fluid \cite{kim2005}. The first terms in eqs. \eqref{r1}-\eqref{r2} are mutual point-like interaction terms and the second terms 
	are 
	self-terms. Notice that it is necessary to take into account the particle radii in the self terms.
	Because of the translational invariance of the system $\bm{G}$ depends only on the relative position $\mathbf{d}$. We are interested in the relative motion, which satisfies  
	the following ODE 
	%
	\begin{eqnarray}
		\hspace{-.5cm} \dot{\mathbf{d}} &=& \frac{1}{8 \pi \mu} \bigg(-\frac{2 k q_1 q_2}{| \mathbf{d} |^3} \,\bm{G}\cdot
		{\mathbf{d}}
		+ (m_2-m_1)g \,\bm{G}\cdot\hat{\mathbf{z}} \bigg) \nonumber \\
				 &-&\!\frac{1}{6\pi\mu} \bigg(\!-\!\frac{k q_1 q_2}{| \mathbf{d} |^3}(\frac{1}{a_1}\!+\!\frac{1}{a_2})
				 {\mathbf{d}}
				 \!+\!(\frac{m_2}{a_2}\!-\!\frac{m_1}{a_1})g \hat{\mathbf{z}} \bigg) \label{eq:masterEquation}
	\end{eqnarray}
%

	Before we examine the properties of 
	\eqref{eq:masterEquation}, 
	we describe physical properties of the system using 
	 non-dimensional parameters which are independent of each other and constant during particle motion
	\begin{eqnarray}
		\gamma = \frac{a_1}{a_2}, \hspace{1.1cm}
		\delta = \frac{m_1}{m_2}, \hspace{1.1cm}
		\beta = -\frac{k q_1 q_2}{L^2 m_2 g}
	\end{eqnarray}
	\noindent so that $\gamma$ is the ratio of particle radii, $\delta$ is the ratio of reduced particle masses and $\beta$ is the ratio of characteristic Coulomb force $F_e=-{k q_1 q_2}/{L^2}$ 
	to characteristic gravitational force $F_g = m_2 g$. 
	The sign of $F_e$ is chosen to be positive when the charges attract each other. There are some physically interesting functions of these parameters. For instance, the ratio of 
	reduced densities is $\delta / \gamma^3$ and the ratio of 
	Stokes velocities is $\delta / \gamma$.
	
	We 
	now choose the units
	\begin{eqnarray}
		L   = a_1+a_2, \hspace{2.1cm}
		V   = \frac{m_2 g}{6\pi\mu L}&&\hspace{0.5cm}
	\end{eqnarray}
	\noindent where $L$ - the characteristic length - is the distance 
	the particle centers would have if the particle surfaces were in contact, and $V$ is a characteristic velocity. These scales define a characteristic time scale $T = {L}/{V}$. 
%
%
	
	%
	Finally, we 
	nondimensionalize the relative position 
	\begin{equation}
		\bm{\alpha} = \frac{\mathbf{d}}{L} 
	\end{equation}
	\noindent 
	so that if the particle surfaces were  
	in contact, $| \bm{\alpha} | = 1$. 
	We can now write equation \eqref{eq:masterEquation} 
	involving only the nondimensional ratios
	\begin{eqnarray}
		\dot{\bm{\alpha}} &=& \frac{3}{2| \bm{\alpha} |^3} \beta \,\bm{\mathcal{G}} \cdot \bm{\alpha} + \frac{3}{4}(1-\delta)\,\bm{\mathcal{G}}\cdot\hat{\mathbf{z}}-\beta\frac{(1+\gamma)^2}{\gamma | \bm{\alpha} |^3} \bm{\alpha} \nonumber \\
				  &-& \frac{(\gamma-\delta)(1+\gamma)}{\gamma} \hat{\mathbf{z}} \label{eq:masterEquationND}
	\end{eqnarray}
 where $\mathcal{G}_{ij} \!= \!\delta_{ij} / | \bm{\alpha} | + \alpha_i \alpha_j / |\bm{\alpha} |^3$ 
 and from now on the dot 
denotes derivative 
with respect to nondimensional time ratio $t/T$.

 
We now analyze eq. \eqref{eq:masterEquationND} and discover a class of vertical configurations which are stable to any perturbation. 

	
	We denote a non-dimensional stationary configuration by $\bm{\alpha}^*
	= \alpha^* \hat{\mathbf{z}}$, with $\alpha^*>0$. Our convention is then to assign label 2 to the particle with larger $\hat{z}$ component in the steady state. To examine the stability of such a configuration, we 
	investigate how the system evolves if we have a first order perturbation $\epsilon$ in the direction perpendicular to gravity and a positive component $\alpha$ in the $\hat{z}$ direction (not necessarily close 
	to $\alpha^*$). If $\bm{\alpha} = \epsilon \hat{\mathbf{x}} + \alpha_z \hat{\mathbf{z}}$ and we neglect second and higher order terms in $\epsilon$ then $\alpha = |\bm{\alpha}|= \sqrt{\alpha_z^2+\epsilon^2}\approx \alpha_z$. 
	With this \eqref{eq:masterEquationND} becomes
	\begin{eqnarray}
		\dot{\epsilon} &=& g(\alpha) \epsilon \label{eq:gForm}\\
		\dot{\alpha} &=& f(\alpha) \label{eq:fForm}
	\end{eqnarray}
	\noindent where 
	\begin{eqnarray}
		g(\alpha)\!\!&=& \!\!\frac{12\gamma\beta - 4(1+\gamma)^2 \beta \alpha + 3\gamma(1-\delta)\alpha^2}{4\gamma\alpha^4} \label{g}\\
		f(\alpha) \!\! &=& \!\! \frac{6\gamma\beta \!-\! 2\beta(1\!+\!\gamma)^2\alpha \!+\! 3\gamma(1\!-\!\delta)\alpha^2 \!-\! 2(\gamma\!-\!\delta)(1\!+\!\gamma)\alpha^3 }{2\gamma\alpha^3} \nonumber \\ \label{f}
	\end{eqnarray}
	Looking at the numerator of $f$, one sees that the first term is the contribution of the mutual electrostatic force, the second term is the contribution of the self electrostatic forces, the third term is the contribution of the mutual gravitational force and the fourth term is the contribution of self gravitational force. Similarly for \eqref{g}.

	For any system of differential equations of the form \eqref{eq:gForm} and \eqref{eq:fForm}, if $g$ and $f$ are continuous then the condition for $\bm{\alpha}^*=\alpha^*\hat{\mathbf{z}}$ 
	to be an steady state is
	\begin{equation}
		f(\alpha^*) = 0 \label{eq:condOne}
	\end{equation}
	\noindent If $f$ is continuously differentiable and $g$ is continuous in an open neighborhood of a steady state $\bm{\alpha}^*$,  then $\bm{\alpha}^*$ is stable if and only if
	\begin{eqnarray}
		g(\alpha^*) &<& 0 \label{eq:condTwo} \\
		f'(\alpha^*) &<& 0. \label{eq:condThree}
	\end{eqnarray}
A proof that \eqref{eq:condOne} - \eqref{eq:condThree} are necessary and sufficient for local asymptotic stability \cite{glendinning1994} is given in section 2 of the Supplemental Material.


	Finally, we impose the feasibility condition
	\begin{equation}
		1 < \alpha^* \label{eq:condFour}
	\end{equation}
	in order to rule out ghostlike overlapping particles. 
	
\begin{figure}[h!]
	\vspace{-0.4cm}
	\hspace{-2cm}\includegraphics[width=0.6\textwidth]{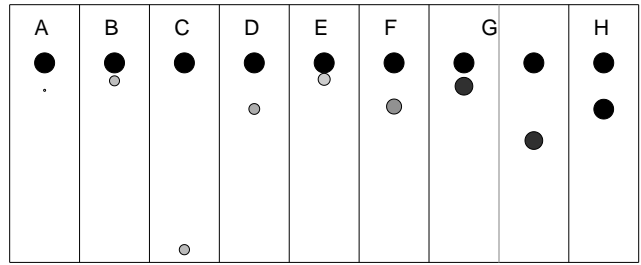}\\
	\vspace{-0.8cm}
	\caption{
	Examples A-H illustrate stable stationary configurations of charged particles settling under gravity in a Stokes flow, for the parameter values listed in Table \ref{t1}. Gravity points down. }\vspace{-0cm}
	\label{fig:particles}
	\end{figure}

	We now demonstrate that there exist solutions to \eqref{eq:condOne} - \eqref{eq:condFour}. We provide examples of stable stationary feasible configurations in figure \ref{fig:particles} with parameters in Table \ref{t1}. 
	In figure \ref{fig:particles}, the density of particle 2 is held constant and painted black, while brighter colors are used to represent denser particles. Similarly, radius $a_2$ of the upper particle is taken to be the same across columns and the radius $a_1$ of the lower particle is drawn to scale. 
%
\begin{table}[h]
\centering
\caption{Positions and parameters of the stable stationary configurations shown in Fig.~\ref{fig:particles}.}\label{t1}
\begin{tabular}{|l|l|l|l|l|l|l|}
\hline
  & $\alpha^*$                                                  & $\beta$    & $\delta$ & $\gamma$ & $\delta / \gamma$ & $\delta / \gamma^3$ \\ \hline
A & 2.5                                                         & 0.160...   & 0.075    & 0.1      & 0.75              & 75                  \\ \hline
B & 1.2                                                         & 0.45       & 0.5      & 0.5      & 1                 &  4                  \\ \hline
C & 12.4                                                        & 2.18...    & 0.5      & 0.51     & 0.980...          &  3.76...          \\ \hline
D & 3                                                           & 0.361...   & 0.5      & 0.54     & 0.925...          &  3.17...          \\ \hline
E & 1.03                                                        & 0.930...   & 1.1      & 0.6      & 1.83...           &  5.09...          \\ \hline
F & 2.5                                                         & 0.523...   & 1        & 0.75     & 1.33...           &  2.37...          \\ \hline
G & \begin{tabular}[c]{@{}l@{}}1.24...\\ 4.13...\end{tabular}   & 0.125      & 0.875    & 0.885    & 0.988...          &  1.26...          \\ \hline
H & 2.33...                                                     & 0.00997... & 0.986    & 0.988    & 0.998...          &  1.02...          \\ \hline
\end{tabular}
\vspace{-.cm}
\end{table}
%

In case A, small $\delta$ and $\gamma$ are chosen. This corresponds to the higher particle being much larger and more massive than the lower particle. Case B shows that stability is possible when $\delta=\gamma$. This corresponds to particles which have identical Stokes velocities. Next we look at cases C \& D where the separation distance $\alpha^*$ is large. Cases E \& F give examples where $\delta / \gamma > 1$, so that the lower particle has a greater Stokes velocity than the upper particle. Case G illustrates that for the same parameters, 
two distinct stable stationary configurations can exist.
In case H, $\gamma \approx 1$, $\delta \approx 1$ and $\beta \approx 0$ showing that there are stable stationary configurations very close to the classic case of two identical uncharged particles.

		Now that we have that the solution set is non-empty, we investigate the range of parameters consistent with a stable feasible steady configuration. The range will come directly from the necessary and sufficient conditions \eqref{eq:condOne} - \eqref{eq:condFour}. The physical implications of these bounds will also be discussed.

		We start with the ratio of characteristic electrostatic to characteristic gravitational force $\beta$. By manipulating the conditions \eqref{eq:condOne} - \eqref{eq:condFour}, one can see

		\begin{equation}
			\frac{3\beta}{\alpha^{*3}} = 3 f(\alpha^*) - \alpha^* (f'(\alpha^*) + 2g(\alpha^*)) > 0 \label{eq:betaBound}
		\end{equation}

		Therefore, if a solution exists, then $\beta >\nobreak0$. This means that the particles must 
		attract each other in order for the system to be stable, in agreement with our predictions which motivated this Letter. This is also important because it allows for stable systems which have a zero net charge, $q_1+q_2=0$. 

		Next, we show that the ratio of reduced masses $\delta >0$. We use that $f(\alpha^*)\! - \!\alpha^* g(\alpha^*)\! >\!0$ to solve for a bound on $\delta$ to get

		\begin{equation}
			\delta > \frac{3\gamma-4\gamma(1+\gamma)\alpha^*}{3\gamma-4(1+\gamma)\alpha^*} > 0 \label{eq:deltaBound}
		\end{equation}
		because the denominator and numerator are both necessarily negative if $\gamma > 0$ and $\alpha^* >1$. 
		This demonstrates that if $m_2 > 0$, then $m_1 > 0$. In the Supplemental Material, we extend this to show that stable doublets can exist only in the $m_2 > 0$ \& $m_1 > 0$ case and the symmetric case when buoyancy dominates over gravity $m_2 < 0$ \& $m_1 < 0$. 


		Moreover, the upper particle must have a larger radius than the lower particle

		\begin{equation}
		 \gamma < 1
		\end{equation}

		The demonstration is somewhat tedious, so it is given in the Supplemental Material.

		If we divide both sides of \eqref{eq:deltaBound} by $\gamma^3$, we can use $\gamma < 1$ and \eqref{eq:condFour} to show the middle term in \eqref{eq:deltaBound} will be larger than one. Therefore, 

		\begin{equation}
		\delta/\gamma^3 > \frac{3-4(1+\gamma)\alpha^*}{(3\gamma-4(1+\gamma)\alpha^*)\gamma^2} > 1
		\end{equation}

		This means that the lower particle has to be more dense than the upper one. This has the interesting implication that in our model stable doublets only form between particles of different material. 


		\begin{figure}
		\includegraphics[width=.55\textwidth]{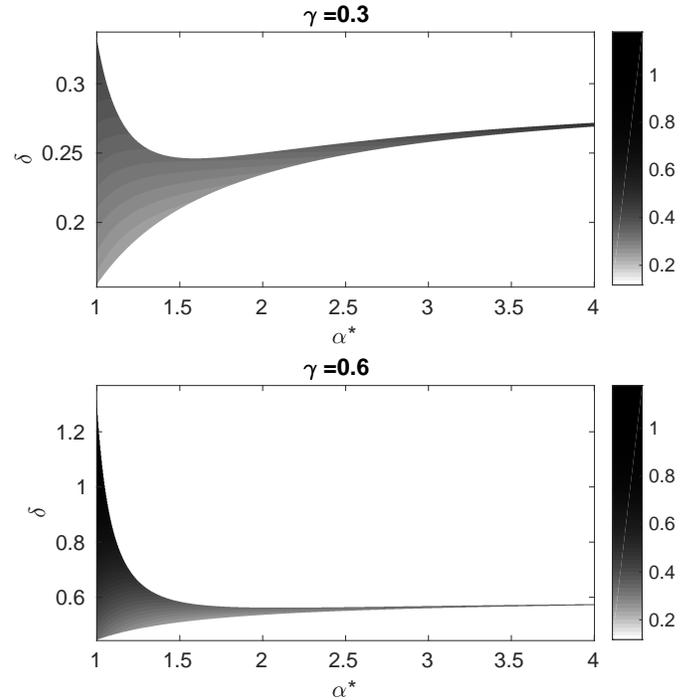}
		\vspace{-0.45cm}
		\caption{Regions of stable steady states in parameter space are plotted. Given $\gamma$ \& $\delta$, the shade at a point is chosen by characteristic force ratio $\beta$ necessary to stabilize the system at $\alpha^*$. If no amount of charge would stabilize a system with the given parameters, the point is left white.
		}\vspace{-0.2cm}
		\label{fig:figures}
		\end{figure}

		With these bounds in mind, we give figure \ref{fig:figures} to illustrate the way $\alpha^*$ and the parameters $\delta$, $\gamma$ and $\beta$ are interrelated. 
		These figures visually demonstrate that the set of parameters that allow a feasible stable steady state is large. 
		One can also see that there exist stable stationary configurations in the "tail" where $\alpha^*$ gets large. 
		Examination of this tail introduces some facts of physical interest. By expanding the relations \eqref{eq:condOne}-\eqref{eq:condFour} in powers of $\frac{1}{\alpha^*}$, we deduce that in the tail the upper particle must have a slightly greater Stokes velocity than the lower one: $1-\frac{\delta}{\gamma} \approx \frac{3(1-\gamma)}{4(1+\gamma)} \frac{1}{\alpha^*}\ll 1$. Looking at the ratio of forces, we see that $\beta \approx \frac{3\gamma (1-\gamma)}{4(1+\gamma)^2} \alpha^* \gg 1$ in the tail. This means that in the tail electrostatic interactions are strong relative to gravitational force. This demonstrates how electrostatic forces can stabilize a doublet even when the distances involved are large.

		In another limit, we keep $\alpha^*$ constant 
		and move values of $\delta$ and $\gamma$ closer and closer to unity. In this limit, the ratio of Stokes velocities $\frac{\delta}{\gamma}$ and relative densities $\frac{\delta}{\gamma^3}$ approach $1$ - that is, the particles get more similar. As a consequence of \eqref{eq:condOne}, 
		$\beta$ scales down to $\beta \ll 1$. We are seeing therefore that a small charge can be expected to stabilize the system in this limit.
		
	In order to aid physical intuition in interpreting the above results, we will conclude by discussing simple examples that demonstrate physically the role of charge in stabilizing a system of settling particles. We will compare system H given in table \ref{t1} and illustrated in figure \ref{fig:particles} with system H', which has the same mass ratio $\delta$ and ratio of radii $\gamma$ but no charge (i.e. $\beta = 0$).

	We start with the uncharged system H'. There is no asymptotically stable $\alpha^*$, so we will instead call a distance ratio $\alpha$' "semi-stable" if it is stable to perturbations in the vertical direction. For an uncharged system, the analog of equation \eqref{eq:condOne} is $3\gamma(1\!-\!\delta) \!-\! 2(\gamma\!-\!\delta)(1\!+\!\gamma)\alpha^* = 0$. The first term is the contribution of the mutual gravitational force and the second term is the contribution of the self gravitational force. The semi-stability condition - the analog of \eqref{eq:condThree} - can be combined with the analog of \eqref{eq:condOne} to get $-2(\gamma\!-\!\delta)(1\!+\!\gamma)<0$. Therefore the semi-stability condition entails ratio of Stokes velocities $\frac{\delta}{\gamma}<1$, so that the self-term  tends to bring the particles together. We also have that $\delta < 1$, therefore the mutual term must tend to push the particles apart. These contributions to the velocity balance exactly at $\alpha$' $\approx 5.21...$. However, if the particles are perturbed even slightly in the horizontal direction the story is different. The analog of \eqref{eq:condThree} is $3\gamma(1-\delta) < 0$, which cannot be satisfied if $\delta < 1$. The horizontal velocity - which contains only this mutual term - is tending to push the particles apart. System H' is not stable.

	Now consider adding a very slight charge to the system so that $\beta = 0.00997...$, in other words system H. In a vertical arrangement, the electrostatic force adds to the self and mutual contributions to velocity without changing their signs, so that the first thing we find is the stationary configuration of the system contracts to $\alpha^* \approx 2.333...$. Now consider horizontal perturbations. There are two electrostatic contributions to the horizontal motion. A mutual term, $12\gamma\beta = 0.118...$, tending to push the particles apart and a self term, $- 4(1+\gamma)^2 \beta \alpha^* = -0.367...$, tending to bring them back. As before, the gravitational contribution to horizontal velocity, $3\gamma(1-\delta)\alpha^2 = 0.225...$, tends to push the particles apart. The restoring term dominates. Therefore, system H is stable. In this way, we have demonstrated how it is possible that qualitatively new behavior can exist in settling charged particle systems, even  with small charge.

	The results in this Letter point toward new fundamental research in the role of charge in the stabilization of semi-dilute polydisperse suspensions of  micro-particles. The core prediction of the model presented in this Letter is the formation of stable asymmetric doublets. The existence of such doublets is experimentally testable. These doublets are not stable without charge, indicating the novelty of the settling dynamics explored here.
\begin{acknowledgments}
This work was supported in part by 
Narodowe Centrum Nauki under grant No.~2014/15/B/ST8/04359. We acknowledge scientific benefits from COST Action MP1305.
\end{acknowledgments}

%

\end{document}